\documentclass[conference]{IEEEtran}
\IEEEoverridecommandlockouts
\usepackage{cite}
\usepackage{amsmath,amssymb,amsfonts}

\usepackage{graphicx}
\usepackage{textcomp}
\usepackage{xcolor}
\def\BibTeX{{\rm B\kern-.05em{\sc i\kern-.025em b}\kern-.08em
    T\kern-.1667em\lower.7ex\hbox{E}\kern-.125emX}}
\usepackage{float}
\graphicspath{ {./images/} }
\usepackage{enumerate}
\usepackage{comment}
\usepackage[hyphens]{url}
\usepackage{soul}
\usepackage[english]{babel}
\usepackage{xspace}
\usepackage{siunitx}
\usepackage[numbers]{natbib}
\usepackage{subfigure}
\usepackage{caption}
\usepackage{subcaption}
\usepackage{algorithm}
\usepackage{algpseudocode}
\usepackage{pifont}
\usepackage{graphicx}
\usepackage{enumitem}
\usepackage{stfloats}
\usepackage{placeins}
\usepackage{stmaryrd} 

\newcommand{\Enc}[1]{\llbracket #1 \rrbracket} 
   
\newcommand{\Sys}{{FIDELIS}\xspace}

\graphicspath{{figures/}} 

\makeatletter
\newcommand{\linebreakand}{%
  \end{@IEEEauthorhalign}
  \hfill\mbox{}\par
  \mbox{}\hfill\begin{@IEEEauthorhalign}
}
\makeatother

\begin{document}


\title{\Sys: Blockchain-Enabled Protection Against Poisoning Attacks in Federated Learning}


\author{
\IEEEauthorblockN{Jane Carney*\thanks{*These authors contributed equally.}}
\IEEEauthorblockA{\textit{Department of Computer Science} \\
\textit{Saint Mary's College of California} \\
jmc61@stmarys-ca.edu}
\and
\IEEEauthorblockN{Kushal Upreti*}
\IEEEauthorblockA{\textit{Department of Computer Science} \\
\textit{North Carolina State University}\\
kupreti@ncsu.edu}

\linebreakand
\IEEEauthorblockN{Gaby G. Dagher} 
\IEEEauthorblockA{\textit{Department of Computer Science} \\
\textit{Boise State University}\\
gabydagher@boisestate.edu}

\and
\IEEEauthorblockN{Tim Andersen}
\IEEEauthorblockA{\textit{Department of Computer Science} \\
\textit{Boise State University}\\
tandersen@boisestate.edu}

}

\maketitle 

\begin{abstract}
Federated learning enhances traditional deep learning by enabling the joint training of a model with the use of IoT device's private data. It ensures privacy for clients, but is susceptible to data poisoning attacks during training that degrade model performance and integrity. Current poisoning detection methods in federated learning lack a standardized detection method or take significant liberties with trust. In this paper, we present \Sys, a novel blockchain-enabled poison detection framework in federated learning. The framework decentralizes the role of the global server across participating clients. We introduce a judge model used to detect data poisoning in model updates. The judge model is produced by each client and verified to reach consensus on a single judge model. We implement our solution to show \Sys is robust against data poisoning attacks and the creation of our judge model is scalable.
\end{abstract}


\section{\uppercase{Introduction}}\label{sec:introduction}
Deep learning has advanced the field of machine learning by providing an in-depth analysis of neural networks for artificial intelligence. In traditional deep learning, a neural network is trained on a large dataset to mimic the brain's decision-making process by analyzing numerous layers in the network. However, this approach raises concerns about the computational power required to train a model and the privacy of the data involved. This leads developers to often outsource the training to a third party. Federated learning (FL) addresses these issue by distributing the operation across numerous nodes, clients, who train the model on their own private data, before sending the model's parameters back to the global server for aggregation. The integration of the Internet of Things (IoT) with FL enables the collection of large amounts of data from devices that can be used for training models. Although federated learning manages data privacy concerns, it is vulnerable to malicious clients who launch data poisoning attacks, diminishing the integrity of the global model~\cite{tolpegin2020data}.

Federated learning enables a distributed machine learning process in which numerous nodes work with a global server to jointly train a model. The global server begins by initializing the global model that is distributed to the nodes. Each node then trains the model locally on their own private data before sending the updated model parameters back to the global server. To preserve privacy, no raw data is shared with the server. The global server aggregates the model parameters using an aggregation formula such as FedAvg~\cite{mcmahan2017communication}, which is the most widely used aggregation formula that calculates the weighted average of all model updates. These steps constitute a training round that is repeated iteratively until the model is fully trained. 

During the federated learning process, malicious clients have the opportunity to launch data poisoning attacks. Data poisoning is harder to detect in a FL environment because the training data utilized is private~\cite{tolpegin2020data}. These attacks aim to influence a model's decision-making process, ultimately affecting its overall performance and behaviors. For example, an attacker may introduce poisoned data in the training of an autonomous vehicle to cause it to misclassify a stop sign and proceed through an intersection. Data poisoning attacks are primarily image-based, but could also include text, speech, and other forms of data. In this paper, we focus on image-based data poisoning through label-flipping.
Label-flipping can be a targeted or an untargeted attack~\cite{lavaur2024systematic}. Targeted attacks~\cite{fan2022survey} are among the most common forms of data poisoning in which an adversary attempts to manipulate a model's behavior for a specific situation. Untargeted attacks~\cite{al2023untargeted} are when an attacker attempts to manipulate training data to degrade the overall performance of the model. Label-flipping attacks occur when an attacker flips the labels of training data to cause misclassifications, e.g., flipping ``spam" emails to ``non-spam" or a ``husky" image to a ``beagle" image. This degrades the model's overall accuracy and performance, making it crucial to have a reliable poisoning detection method.

Federated learning relies on a global server to distribute the global model and aggregate updates from clients. The server may also run poisoning detection methods to find malicious updates. This places a significant amount of trust in the server's ability to perform these tasks accurately. To reduce this dependence on a single entity, distributed approaches should be considered.

Current strategies used to detect poisoning in federated learning take significant liberties with trust or lack a unified detection approach. Some solutions depend on a trusted entity to perform key aspects of their detection method~\cite{raza2022using,xie2019zeno}. This is an unrealistic expectation when the focus is finding untrustworthy clients. Other methods utilize blockchain technology~\cite{weng2019deepchain,li2021blockchain,desai2021blockfla} in FL to improve security and trust among participants, but lack a clear and standardized method to detect poisoned models. This highlights the need for a secure solution to detect poisoning attacks in federated learning and maintain overall model integrity. 

In this paper, we propose our framework \Sys that leverages blockchain technology to decentralize the federated learning environment and create a judge model used for poison detection. The judge model is reached through consensus on the blockchain between clients. Model updates that are deemed malicious by the judge model are excluded from aggregation to ensure the global model's integrity.
\subsection{Contributions}

This paper presents multiple contributions to poisoning detection in federated learning via blockchain:
\begin{itemize}
    \setlength\itemsep{0em}
    \item We propose a novel framework, named \Sys, to detect poisoning in federated learning. \Sys utilizes blockchain technology to replace the trusted global server with a decentralized system that distributes the trust among clients. 
    \item \Sys enables clients to reach consensus on a \textit{judge model} to be used to detect data poisoning in model updates. We extract the movement of gradients in benign training to train the judge model.
    \item We conducted a comprehensive experimental evaluation. The results show that the process for creating the judge model is scalable with respect to a linear increase in the number of clients and show that \Sys achieves a high level of accuracy for detecting poisoned model updates.
\end{itemize}
\FloatBarrier
\section{Related Work}\label{sec:related_work}

Significant research has been conducted to address data poisoning attacks in federated learning. This section reviews the main areas of work: blockchain-enabled federated learning frameworks, federated learning frameworks, and comparison mechanisms.

\subsection{Blockchain-Enabled FL}

Blockchain technologies have been utilized to create a distributed network of miners that work to create a secure and immutable ledger. Its decentralized and distributed nature fits well in a federated learning environment. Research has been done in this area to show how blockchain can be used in FL to train a model.

Weng et al. introduced DeepChain~\cite{weng2019deepchain} as a deep learning framework based on blockchain. It uses an incentive mechanism and transactions to promote the joint and honest training of a model through gradient collection and parameter updates. All gradient updates and model aggregation are recorded on the chain with smart contracts. They use DeepCoin as an incentive and penalty token. Similarly, Li et al. proposed BLADE-FL~\cite{li2021blockchain} that uses blockchain to decentralize FL. It uses reputation and incentives to penalize nodes. It focuses on optimization and efficiency. Both approaches, as well as a few others~\cite{feng2021bafl,sarhan2022hbfl} decentralize the FL process and create immutable records of model updates with incentive mechanisms.

Desai et al. proposed BlockFLA~\cite{desai2021blockfla} which provides a general FL framework that any aggregation function and detection algorithm can be incorporated into. It uses a private blockchain for model updates and aggregation, and a public blockchain for cryptocurrency transactions through smart contracts. 

These methods distribute the role of the central server and provide frameworks for incentive mechanisms, consensus protocol, and an immutable ledger. However, they fail to provide a distinct method for poison detection. DeepChain attempts to address it through penalizing tokens that discourage malicious updates. Though, no specific algorithms are presented to detect malicious updates. Our framework addresses this gap by embedding a poisoning detection method within the B-FL process.

\subsection{Federated Learning Detection}

Various poisoning detection methods have been implemented in federated learning to mitigate malicious updates. Some approaches focus on measuring distances between updates using different metrics, such as Euclidean distance~\cite{herath2023recursive}, cosine similarity~\cite{zhu2024byzantine}, and jaccard similarity~\cite{doku2021mitigating}. These methods work by finding updates that deviate significantly from each other; a larger distance infers anomalous behavior. However, adversaries can craft their model updates to avoid detection by doing smaller amounts of poisoning. As a result, distance-based techniques are not always reliable in detecting poisoning.

Robust aggregation methods have been proposed, such as Krum~\cite{blanchard2017machine}, Trimmed Mean~\cite{wang2025federated}, and Median~\cite{pillutla2022robust} to defend against Byzantine and poisoning attacks. These methods attempt to filter out outliers. Krum starts by computing the similarity of the updates using the Euclidean distance and then selects the update with the lowest Krum score as the only update for the training round. The server is used to perform these various aggregation methods. This places full trust in the global server to be able to accurately execute the aggregation methods and assumes that the global server is not malicious. This centralized reliance creates a single point of failure for the FL process. Our proposed blockchain framework addresses this issue by decentralizing the role of the global server.

\subsection{Comparison Mechanisms}

Methods in this category compare updates to a clean dataset or model to detect anomalous behavior. Xie et al. proposed Zeno~\cite{xie2019zeno} which uses a reference batch dataset to compare updates and calculate a maliciousness score. The reference batch is a small, trusted dataset held locally by the sever. However, it is not specified where the reference batch is derived, and this method relies on the server to perform the detection.

Similarly, Raza et al. proposed the creation of a reference model and an auditor model~\cite{raza2022using}. A trusted third party begins by training the reference model on a public dataset that is used to create the audit dataset for the audit model. All model updates are run through the audit model to detect anomalies. If the amount of anomalies detected exceeds a calculated poison threshold, the update is discarded from aggregation. This approach relies entirely on a trusted third party to create both the reference and audit model. This liberty in trust is unrealistic in an untrustworthy environment.

Yang et al. proposed a B-FL architecture for trustworthy AI training~\cite{yang2022trustworthy}. It uses consensus on the blockchain to validate the correctness of updates with a small, trusted validation dataset for verification. The dataset is either derived from a public dataset for the specified FL training task or from edge-collected data. If the dataset is derived from edge collection or a public dataset it may not be representative of the training task or be poisoned itself, making it inaccurate for poison detection. Our proposed judge model used for poison detection is chosen through consensus on the blockchain to mitigate reliance on any single trusted entity.

\begin{table*}[t]
\begin{center}
\label{relatedWorksTable}
\begin{tabular}{|l|c|c|c|c|c|c|}
 \hline
 \multicolumn{1}{|c|}{}&\multicolumn{1}{c}{\textbf{Decentralized}}&\multicolumn{2}{|c|}{\textbf{Attack Focus}}&\multicolumn{1}{c|}{\textbf{Detection}}\\
 \cline{3-4}
  \textbf{Related Works} & \textbf{Global Server} & \textbf{Data Poisoning} & \textbf{Model Poisoning} & \textbf{Method} \\
 \hline
 Raza \textit{et al.}~\cite{raza2022using}& & Label-flipping & Sign Flipping & SVM\\
 \hline
 Yang \textit{et al.}~\cite{yang2022trustworthy}& $\checkmark$ & & Byzantine Attacks & Cross Validation\\
 \hline
 Xie \textit{et al.}~\cite{xie2019zeno}& & Backdoor Attacks & & Cross Validation\\
 \hline
 Desai \textit{et al.}~\cite{desai2021blockfla}& $\checkmark$ & Backdoor Attacks & & Client-defined algorithm\\
 \hline
\hline
 Our Framework: \Sys & $\checkmark$ & Label-flipping &  & Isolation Forrest\\
 \hline
\end{tabular}
\caption{Comparative evaluation of our work with related works in the areas of: Decentralized Global Server, Attack Focus (Data Poisoning and Model Poisoning), and Detection Method.}
\end{center}
\end{table*}

\section{Problem Formulation}\label{sec:problem_formulation}

In this section, we define the problems being addressed in our work and the specific assumptions of the adversary model. This outlines the goals and capabilities of the adversary, as well as the precise problem which we are focused on solving with \Sys.

\subsection{Adversarial Model}
In \Sys there is one main party: the clients. Clients may behave as expected or they may act maliciously. Malicious clients have two main goals. First, poison the judge model so that it performs poorly on anomaly detection, causing the judge model to incorrectly allow poisoned models to be included during aggregation. Second, poison their local models during training via label flipping attacks to degrade the performance of the global model, leading to frequent misclassification of images. We consider the following assumptions of the adversary model:

\begin{itemize}
    \item There are at least one or more malicious clients in the federated learning environment whose objective is to poison the judge model, global model, or both, potentially in a collaborative manner. Malicious clients do not hold a majority of the total number of clients in the system.
    \item All clients, both benign and malicious, must adhere to the standard federated learning protocol and the proposed blockchain protocol. Malicious clients are unable to compromise the training of benign clients.
    \item All clients have access to the public dataset, global model, and the finalized judge model. The public dataset cannot be modified. Once chosen, a judge cannot be changed and will be used throughout the training process. No single client can change the global model directly, and updates to the global model must be reached through consensus between each training round.
    \item The strategy utilized by malicious clients will be label flipping attacks either targeted or untargeted. Additionally, malicious clients have the option to determine the timing of their attacks, i.e., they can choose to refrain from poisoning a model during a given round.
\end{itemize}

\begin{table}[h]
\begin{center}
    \label{tab:my_label}
    \caption{Notation Table}
    \begin{tabular}{|c|c|c|c|}
    \hline
        $c_i$ & Clients & $v_{n,n}$ & Client's votes \\
        \hline
        $D_i$ & Client's Private Data & $V_{HT}$ & Homomorphic total\\
        \hline
        $M_0$ & Original CNN Model & $L_i$ & Judge Model Lists\\
        \hline
        $M_i$ & Trained Model Updates & $S_i$ & Source Model\\
        \hline
        $\hat{P}$ & Public Dataset & $Q_i$ & Accepted Models Lists\\
        \hline
        $J_0$ & Initial Judge Model & $J_i$ & Trained Judge Model\\
        \hline
    \end{tabular}
    \end{center}
\end{table}

\subsection{Problem Statement}

Let $c_1, c_2, \dots, c_n$, where $n > 2$, be the set of clients in the federated learning system. Each client $c_i$ possesses their own private datasets $D_i$ with which they use to train their local models $M_i$ during each round of training. Let $\hat{P}$, $M_0$, and $J_0$, be the public dataset, initial model, and initial judge model, respectively. The objective of our research is to propose a blockchain-based framework which aims to produce a global model, $G$, such that (1) trust is decentralized among clients, (2) consensus is reached on a model used for detecting anomalous $M_i$, and (3) the global model integrity throughout the training process is maintained.


\section{\uppercase{Solution: \Sys}}\label{sec:solution}
\subsection{Solution Overview}

\Sys introduces a poisoning detection method for a neural network model. By decentralizing the federated learning process with blockchain, our approach removes the need for a trusted central authority. The main focus of the framework is to mitigate malicious model updates to improve the overall integrity of the model. An isolation forest is used to create a judge model to detect poisoning before aggregation. The proposed framework verifies the work of clients on the blockchain to produce a reliable model.

An applicant submits a request for model training. In their request, they must provide the initial model, isolation forest, and a public dataset specific to the training needed. Clients on the blockchain begin a training round by producing a judge model for poison detection and a trained local model. The judge model is created using the public dataset and the isolation forest. Clients train the initial model locally with their own private data to produce model updates. After training, clients submit these two models to the blockchain. 

\begin{figure*} [t]
    \centering
    \includegraphics[width=1\linewidth]{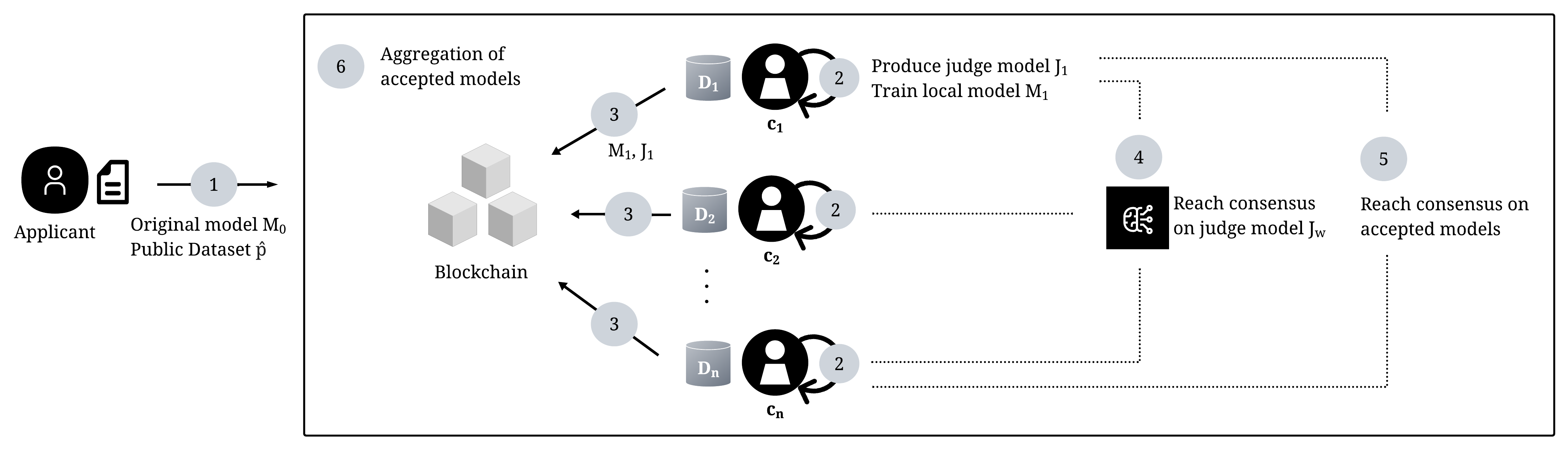}
    \caption{Overview of \Sys.}
    \label{fig:Overview}
\end{figure*}

To reach consensus on a sole judge model, clients test each judge model with the public dataset. This provides them with an anomaly score for each model that allows them to pass, 1, or fail, 0, the judge models. The judge model that receives the highest tally of 1's is deemed the sole judge model. After a judge model is determined, clients test every model update through the judge model. If a model is marked as anomalous, it will not be included in aggregation. All accepted models are aggregated together. Figure-\ref{fig:Overview} illustrates the overall process of our framework \Sys.

\subsection{Local Training of the Client and Judge Models}

Given $n$ clients $c_1, c_2, \dots, c_n$, $ 2 \leq n $, in the federated learning system, each client $c_i$ trains two models: a judge model $J_i$ to be used for poison detection, and a model update $M_i$. By having each client train the judge model, no trusted entity is relied on to provide the poison detection method. 

The judge model is an isolation forest. To train the forest, clients first train the initial model $M_0$ with the public dataset $\hat{P}$ to produce a source model $S_i$. From this source model, clients can gather batch statistics about how the gradients move and change from each batch. There are nine statistics used: mean, standard deviation, minimum, maximum, range, skew, kurtosis, L1, and L2. To account for different batch sizes in data, a five-stat summary is created from the nine statistics. The five-stat summary takes the mean, standard deviation, minimum, maximum, and range of each of the nine statistics. This creates a total of forty-five data points. A client will simulate this process of collecting data points for other clients to reach a sufficient amount of data to train the isolation forest. Algorithm 1 outlines the steps for a client to train a judge model.

\algrenewcommand\algorithmicrequire{\textbf{Input:}}
\algrenewcommand\algorithmicensure{\textbf{Output:}}

\begin{algorithm}
\caption{Training of Judge Model $J_i$}
\begin{algorithmic}[1]
\Require Public Dataset $\hat{P}$, Initial Model $M_0$
\Ensure Judge Model $J_i$
\State $\hat{P}$ is split into $\hat{P}_{train}$ and $\hat{P}_{test}$
\For{$n$ number of simulations}
    \State $\hat{P}_{train}$ trains $M_0$ to produce a source model $S_i$
    \State Gather nine batch statistics
    \State Create a five-stat summary for each statistic to \Statex \hspace{\algorithmicindent} attain forty-five data points 
    \State Append data points to list $T_i$ 
\EndFor
\State Train the initial isolation forest $J_0$ with the 1-D list $T_i$ to produce $J_i$
\State \Return $J_i$
\end{algorithmic}
\end{algorithm}

After producing a judge model, clients train the initial model locally. They use their own private data $P_i$ and the initial model $M_0$ to produce $M_i$. They then submit the model parameters of their trained model $M_i$ and their judge model $J_i$ to the blockchain.

\subsection{Confirmation of Judge Model}

Clients must reach consensus on one universal judge model that will be used for poison detection in all model updates. To begin, each client tests each judge model $J_1, J_2, \dots, J_n$ with $\hat{P}_{test}$. Each judge model will output a 1 to indicate benign or -1 to indicate anomalous. Since $\hat{P}_{test}$ is a clean dataset, a judge model should output a 1. Clients produce a list $L_i$ where they pass, 1, or fail, 0, each judge model depending on whether it marks the dataset anomalous.

Clients encrypt their votes in their list $L_i$ before publishing it on the blockchain to prevent malicious clients from attempting to manipulate the vote. Clients use ElGamal encryption to encrypt their votes with the public group key. They then homomorphically add the corresponding rows from $L_1, L_2, \dots, L_n$ to produce the list $L_{HT}$. They collectively decrypt $L_{HT}$ to obtain the homomorphic total score for each model. Table~\ref{tab:encryption} illustrates this process.

\begin{table} [h]
    \centering
    \begin{tabular}{cccccc}
         & $L_1$ & $L_2$ & \dots & $L_n$ & $L_{HT}$\\
        $J_1$ & $\Enc{v_{1,1}}$ +& $\Enc{v_{1,2}}$ + & \dots & + $\Enc{v_{1,n}}$ =& $\Enc{V_1}$\\
        $J_2$ & $\Enc{v_{2,1}}$ +& $\Enc{v_{2,2}}$ + & \dots & + $\Enc{v_{2,n}}$ =& $\Enc{V_2}$ \\
        \vdots &  &  &  & & \vdots \\
        $J_n$ & $\Enc{v_{n,1}}$ +& $\Enc{v_{n,2}}$ +& \dots &+ $\Enc{v_{n,n}}$ =& $\Enc{V_n}$\\ 
    \end{tabular}
    \caption{Homomorphic tallying of all votes for each judge model.}
    \label{tab:encryption}
\end{table}

\begin{figure*}
    \centering
    \includegraphics[width=1\linewidth]{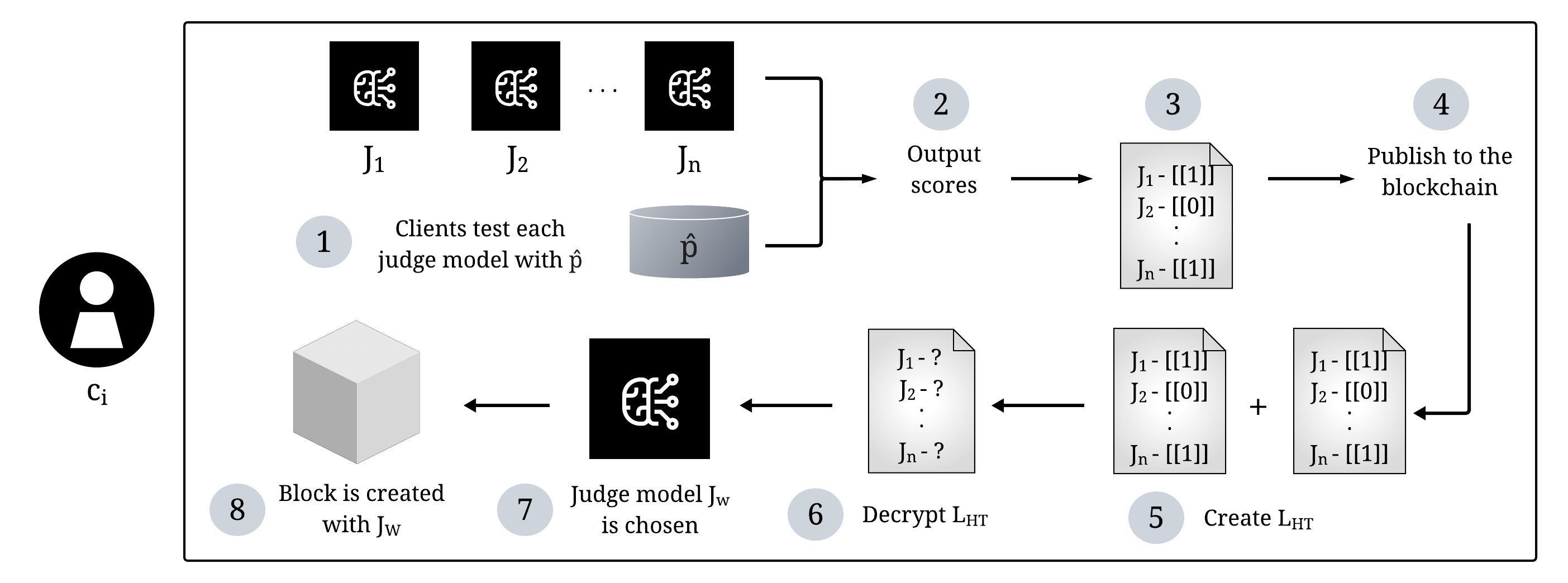}
    \caption{Diagram showing how consensus is reached on a judge model. In step 3, clients pass, 1, or fail, 0, each model and encrypt the values. In step 5, clients homomorphically add corresponding rows to create $L_{HT}$.}
    \label{fig:JudgeConsensus}
\end{figure*}

The judge model with the highest total score is the winning judge model $J_w$ to be used for poison detection in all model updates. This process ensures that the most accurate model is chosen by having all clients test and score all judge models. Figure~\ref{fig:JudgeConsensus} demonstrates the process of reaching consensus on a judge model.

\subsection{Consensus of Accepted Model Updates}

\begin{figure*}
    \centering
    \includegraphics[width=1\linewidth]{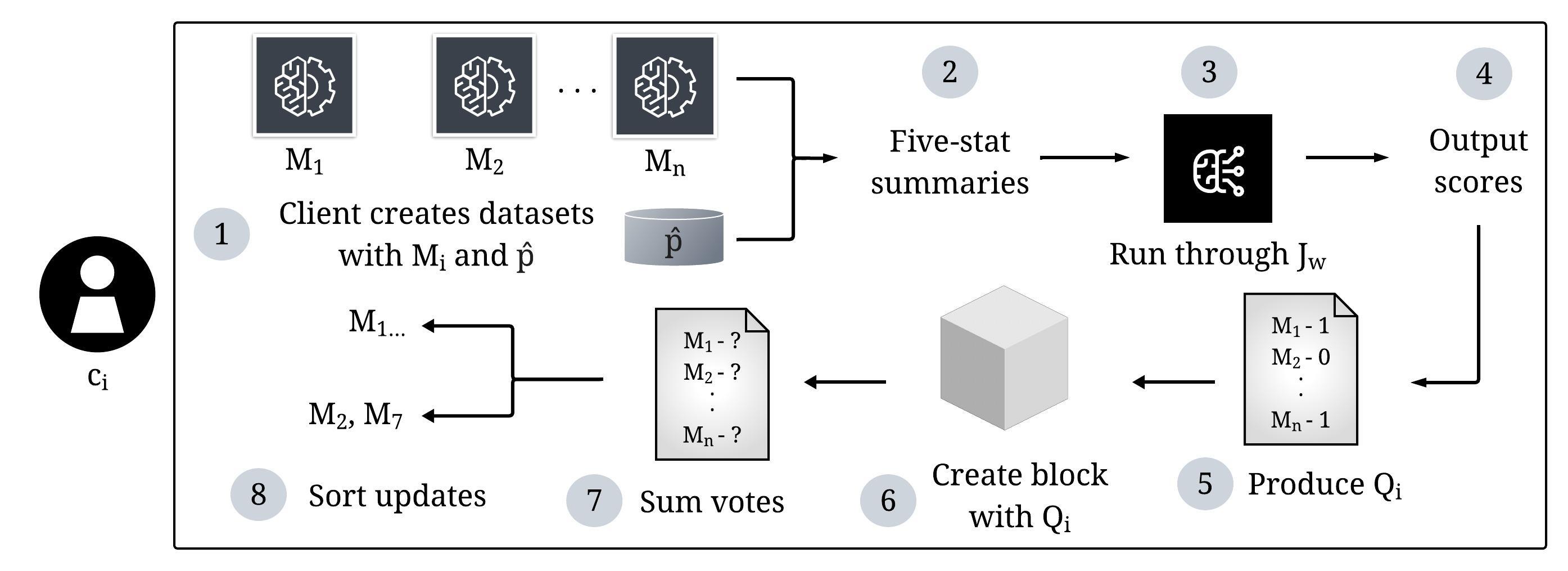}
    \caption{Diagram depicting how consensus is reached on accepted models.}
    \label{fig:acceptedconsensus}
\end{figure*}

Clients must reach consensus on which models should be accepted in aggregation. A model update deemed anomalous will not be included in aggregation. Each client will run all model updates $M_1, M_2, \dots, M_n$ with $\hat{P}_{test}$ to observe the gradient changes without applying them and create five-stat summaries. This allows clients to not update the models during this process. Clients run the five-stat summaries through the judge model. The judge model will mark a model update as 1, benign, or -1 anomalous. Clients then vote to pass, 1, or fail, 0, each model update to create the list $Q_i$. Clients create a block with their list $Q_i$ and put their signature on it. Clients can then total the votes from all blocks to determine which models are accepted. Since we assume malicious clients do not hold a majority in the system, a model update must receive over 50\% of the vote to be accepted in aggregation. All other updates will be discarded. This process ensures malicious model updates are not included in aggregation by having all clients check all updates. Figure~\ref{fig:acceptedconsensus} illustrates this process of reaching on consensus on which model updates will be included in aggregation.

\subsection{Aggregation}

Once all accepted models have been determined, aggregation occurs. We use federated averaging~\cite{mcmahan2017communication} to calculate the weighted average of all accepted updates. FedAvg is given by the following formula:
\begin{equation}
w_{t+1} = \sum_{k=1}^K \frac{n_k}{n} w_t^k
\end{equation}
Where $w_{t+1}$ are the global parameters after aggregation, $w_t^k$ are the model parameters from client $k$ at round $t$, $n_k$ are the number of data points for client $k$, and $K$ are the number of clients participating in a training round.

We also considered weighted aggregation approaches, such as FedLAW~\cite{li2023revisiting} which weights updates based on the similarity of the data distribution and RFA~\cite{pillutla2022robust} that weights using the geometric median to remove outliers. 

\section{Experimental Evaluation}\label{sec:experimental_evaluation}
In this section, we evaluate the performance of \Sys based on different metrics. This includes the accuracy of the global model, the classification accuracy of the judge model, and the scalability of creating the judge model. We implemented \Sys for testing. 

\subsection{Implementation and Setup}
\textbf{Hardware.} All experiments were conducted using two remote machines, each with an Intel Xeon w9-3495X CPU (56 cores, 112 threads, 4.8 GHz), 251 GiB of system RAM, and a NVIDIA RTX 6000 Ada Generation GPU with 48 GiB of VRAM.

\textbf{Dataset.} We used two datasets for our experiments: the Oxford-IIIT Pet Dataset and the Cat vs. Dog Dataset, provided through a partnership between Petfinder and Microsoft. The Oxford-IIIT Dataset contains approximately 7,300 images, representing 25 dog breeds and 12 cat breeds. We used it as the public dataset for creating the judge model during our experiments. The Cat vs. Dog Dataset includes around 25,000 images, evenly split between dogs and cats, and was used to simulate local client data in our federated learning experiments.

\textbf{Preprocessing.} Several transformations were applied to the datasets before and during experimentation. For the Oxford-IIIT Pet Dataset, we relabeled the images into two classes, Dogs and Cats, discarding the original 37 breed categories. Due to class imbalance, we limited the total number of images to match the size of the smaller class, Cats, thus reducing the dataset to approximately 4,800 images. We then split this subset into 90/10 for training and testing. For the Cat vs. Dog Dataset, we reorganized the images to simulate local client data, dividing them for three separate cases: 50, 100, and 150 clients. Each client was assigned only training images, since testing was performed using images from the public dataset. We partitioned the Cat vs. Dog dataset into shared and unique portions using splits of 60/40, 65/35, and 70/30, respectively, where the larger split served as the shared image pool and the smaller split served as the unique image pool. Unique images were evenly distributed among clients, while images from the shared pool were randomly assigned to each client thus allowing for potential overlap. Each client was allocated 400 images per class (800 total). This design choice was done for for two main reasons: data limitations and to reflect a more realistic real-world scenario. Limiting image counts allowed us to maximize our federated
learning environment to represent a more realistic scenario while still maintaining a reasonable distribution between clients. In practice, client datasets are unlikely to be entirely unique, images of cats and dogs often share similar features especially within the same breed, so overlap between image features should be expected. To simulate malicious behavior, we implemented label flipping attacks during training. More specifically, each malicious client poisoned 35$\%$ of their dataset by flipping the labels of dog images to cat.

\textbf{Training Parameters.} Due to the intense computational power requirements and dataset limitations, we utilized the RESNET-18 CNN model preloaded with its default weights. We changed the last fully connected layer to output two classes: dog or cat. We kept all layers unfrozen so that all weights would change during training. For all training, we utilized the Adam optimizer with the Cross Entropy Loss function. We decided to use Adam since we are running our training for a shorter number of epochs than normal, and our dataset is considered relatively small in the realm of machine learning, particularly for image classification. We decided to use the Cross Entropy Loss function as it is generally the standard when it comes to image classification tasks. For all experiments, we decided to do a total of 5 epochs per client, where we aggregated the models between each epoch. The batch size was set to 8, and the learning rate was set to $1 \times 10^{-4}$. We decided to use a low number of epochs due to our baseline model already being pretrained, thus it is likely to converge much faster. Likewise, we decided to use a relatively smaller learning rate as we don’t want the weights to change in a drastic manner.

\subsection{Experiment 1: Judge Model Accuracy}
\begin{figure*} [t]
    \centering
    \subfigure[50 Clients]{\includegraphics[width=0.32\textwidth]{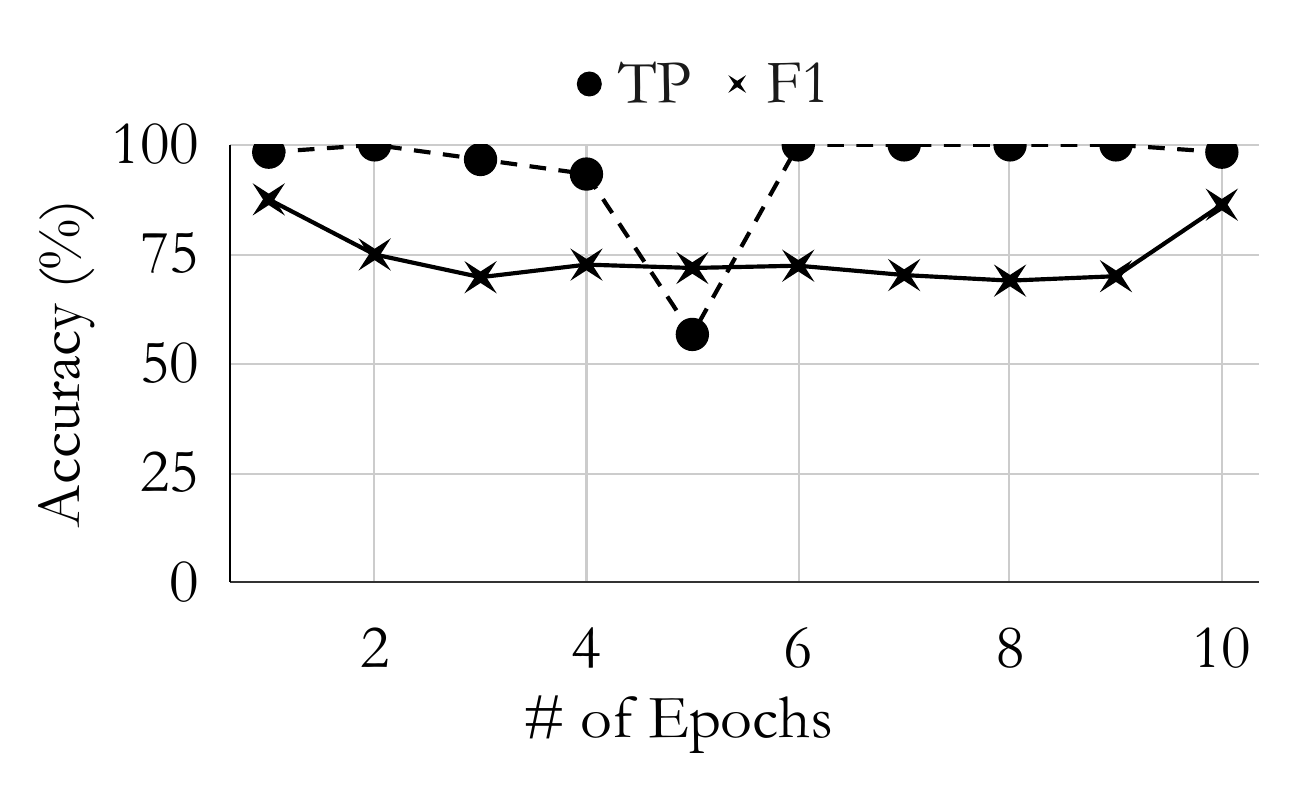}}
    \hfill
    \subfigure[100 Clients]{\includegraphics[width=0.32\textwidth]{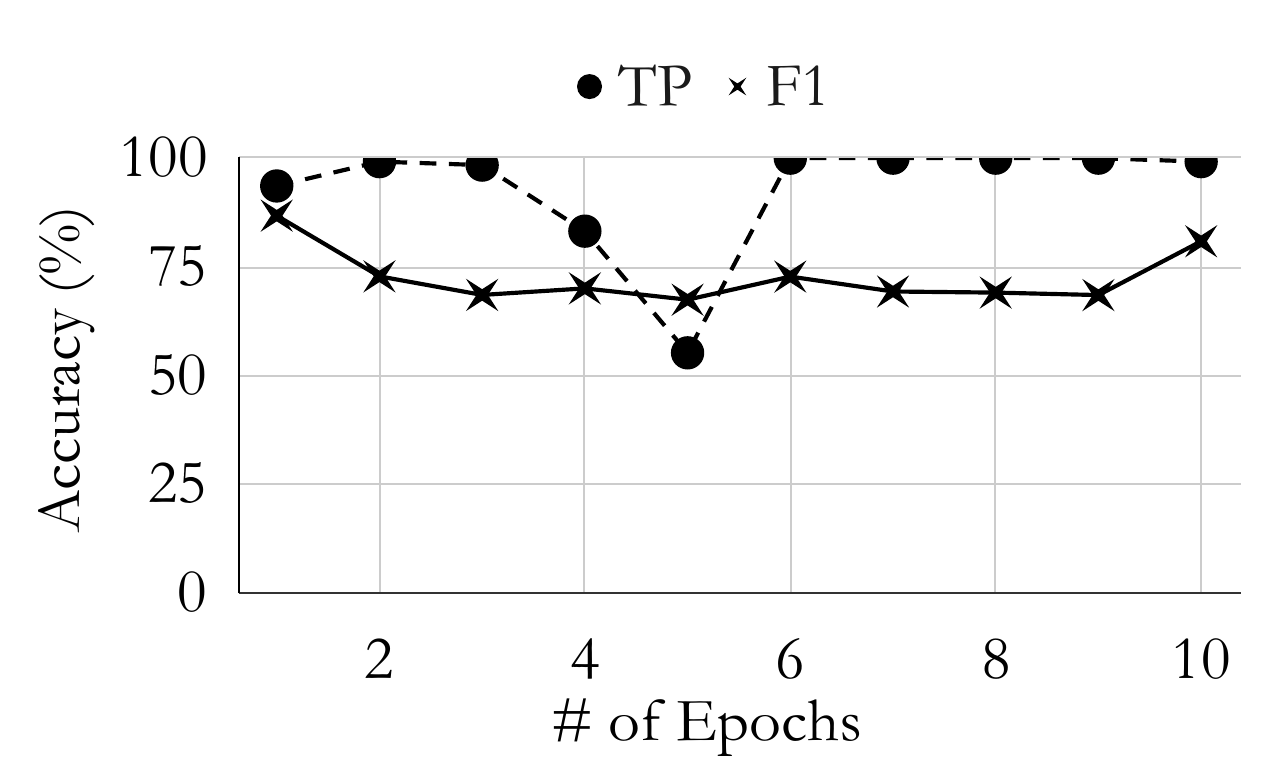}}
    \hfill
    \subfigure[150 Clients]{\includegraphics[width=0.32\textwidth]{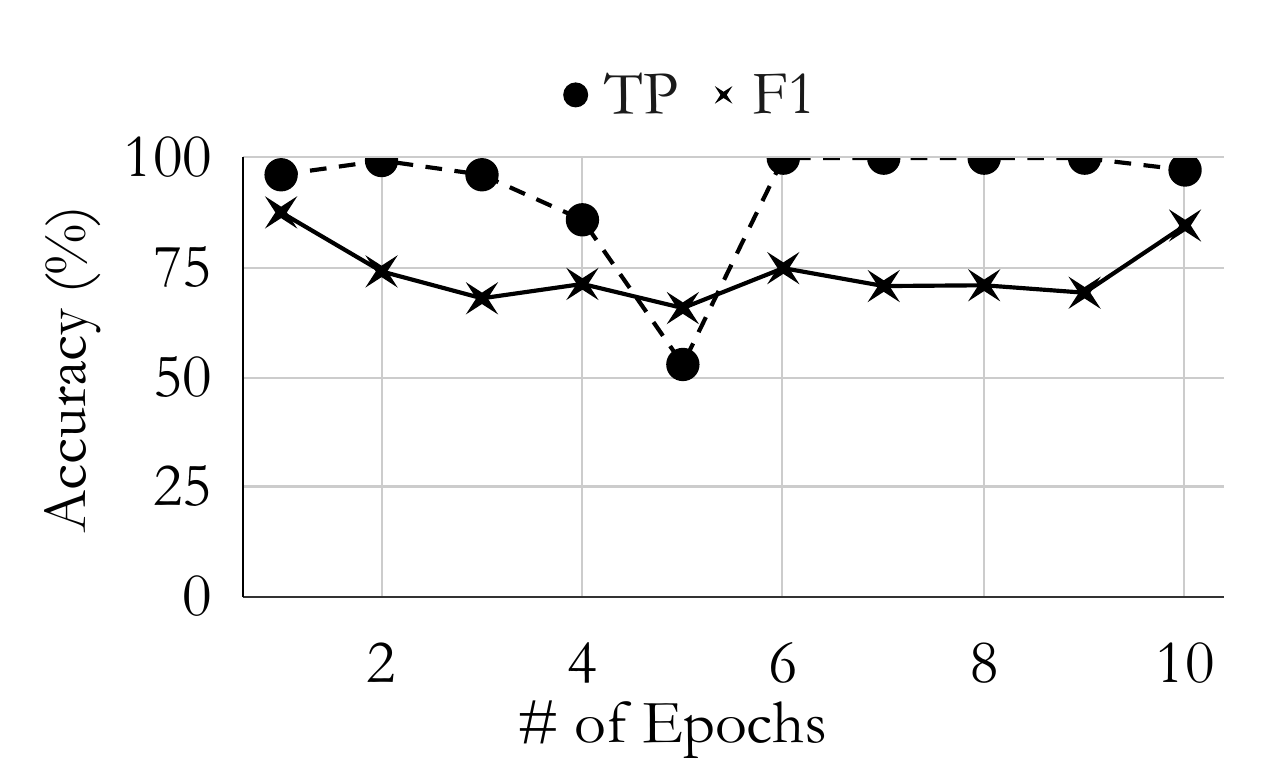}}
    \caption{Predictive Performance of the Judge Model for True Positives and the F1 score for 50, 100, and 150 clients.}
    \label{fig:TPF1}
\end{figure*}

This experiment analyzes the judge model's accuracy in classifying model updates. We capture the true positive rates and F1 score\textemdash{}which depicts accuracy taking into consideration the true positives, false positives, and false negatives. We varied the number of clients from 50, 100, and 150 with 25\% being malicious in each scenario, as shown in Figure~\ref{fig:TPF1}.

 We observe that our judge model has a high rate of malicious update detection when training is run with 6 epochs or more with accuracy floating around 95-100\% excluding the drop during epoch 5. The F1 score remains moderately high throughout the entirety of training, remaining relatively constant around a rate of 70-75\% reaching a max of 90\% during early and late stages of training. Both TP and F1 accuracy stay consistent regardless of how many clients are in the system.

\subsection{Experiment 2: Global Model Accuracy}
\begin{figure} [h]
    \centering
    \includegraphics[width=1\linewidth]{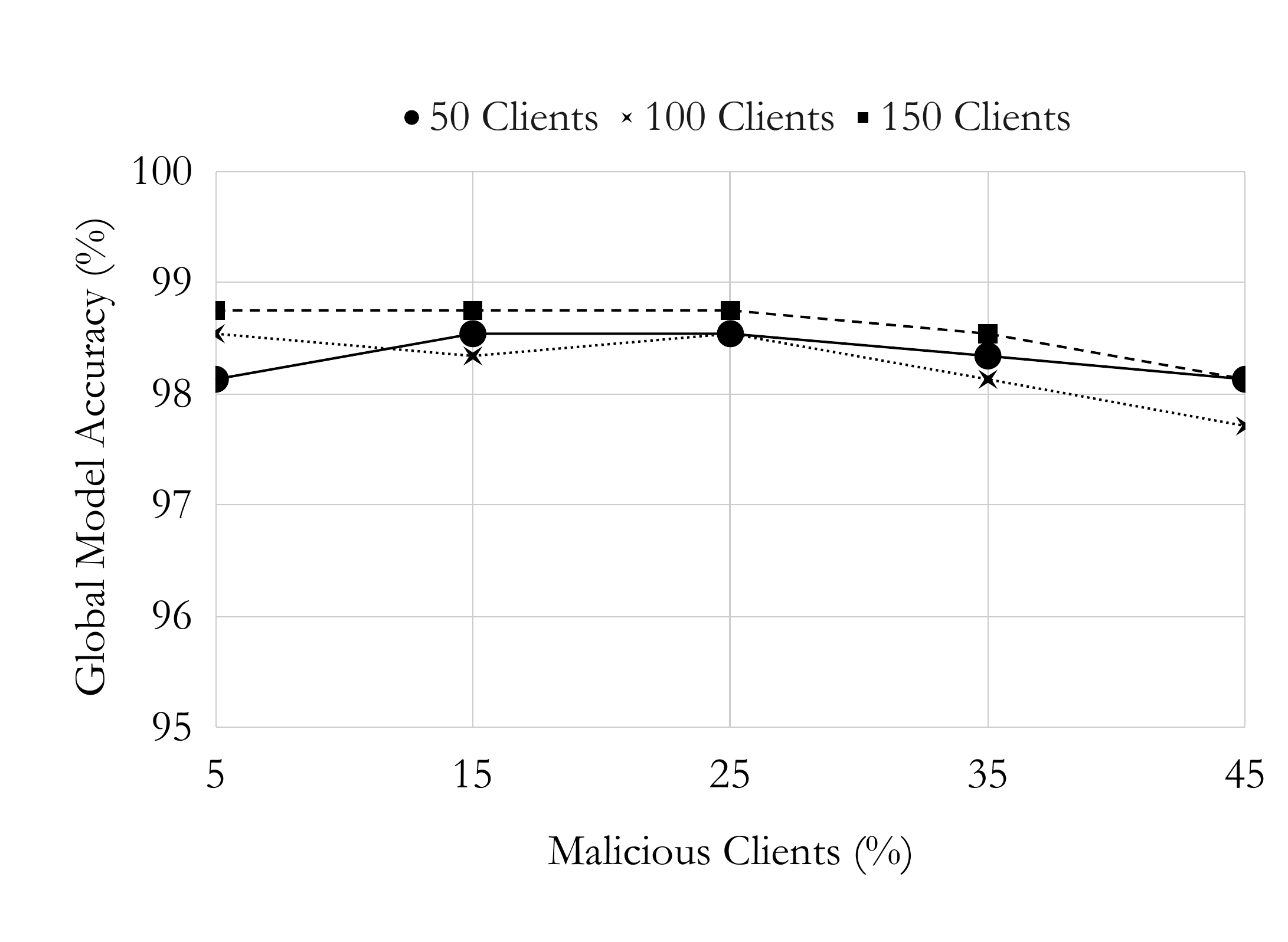}
    \caption{Global Model Accuracy}
    \label{fig:gmaccuracy}
\end{figure}

This experiment measures the accuracy of the global model in the classification of images versus the percentage of malicious clients in the system. The percentage of malicious clients was tested at $5\%$ and increased in increments of $10\%$ up to $45\%$. We tested with varying numbers of clients in the system: 50, 100, and 150. Figure~\ref{fig:gmaccuracy} depicts the global model accuracy as the percentage of malicious clients in the system increases.

We observe that our system's global model accuracy remains above $98\%$ with up to $35\%$ of the system containing malicious clients. Regardless of the number of clients in the system, we conclude that the accuracy remains high, and \Sys is robust with respect to the number of malicious clients up to $35\%$.

\subsection{Experiment 3: Runtime of Judge Model Creation}
\begin{figure} [h]
    \centering
    \includegraphics[width=1\linewidth]{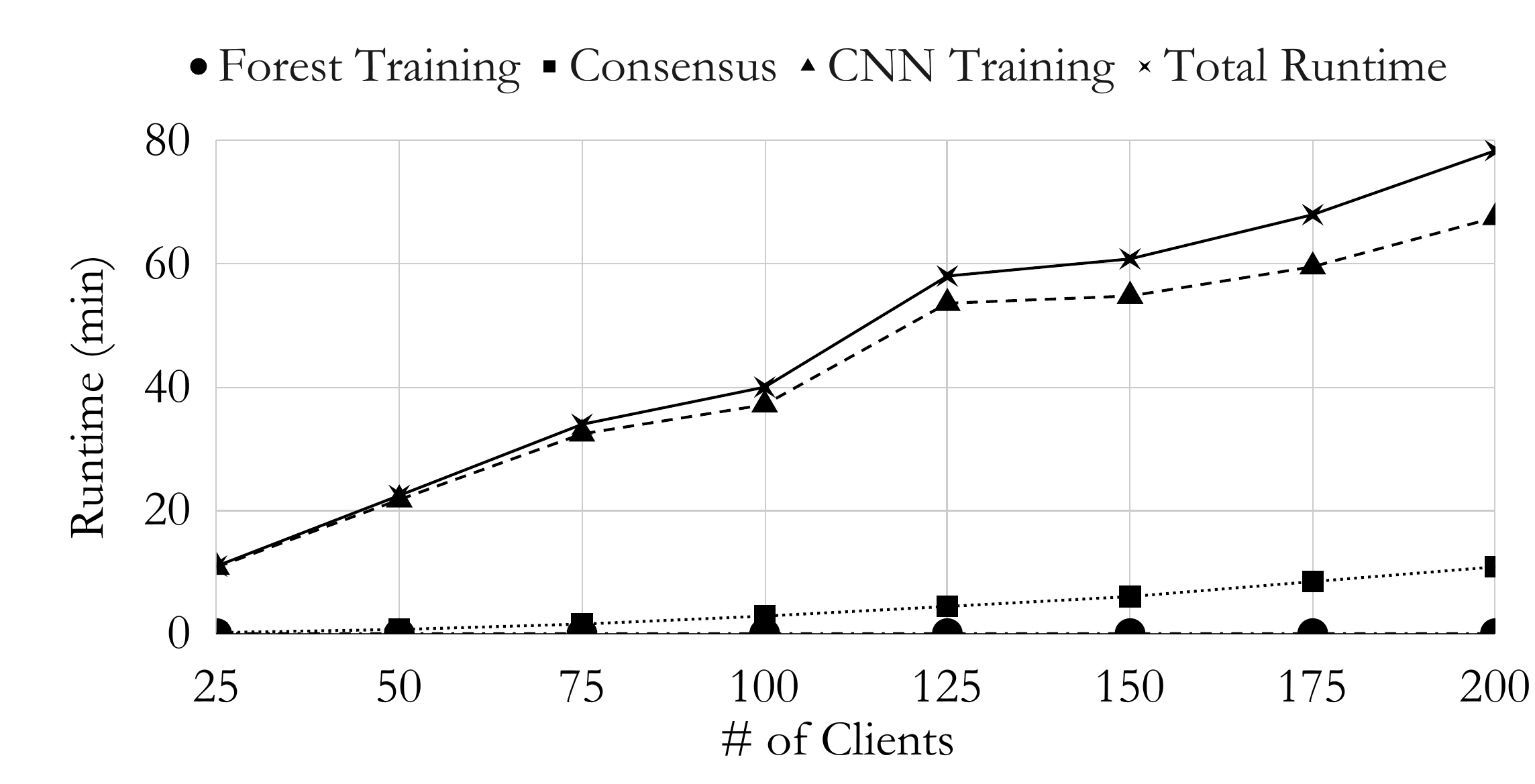}
    \caption{Judge Model Creation Runtime}
    \label{fig:runtime}
\end{figure}

In this experiment, we evaluate the scalability of the process of creating the judge model versus the number of clients in the system. We measure the runtime when the number of clients grow from 25 to 200, increasing in increments of 25. Figure~\ref{fig:runtime} shows the runtime for the creation of the judge model considering the number of clients in the system.

We observe that the overall runtime of the judge model creation increases linearly as the number of clients increases linearly. We observe the majority of runtime is compromised of CNN training, whereas the isolation forest training is the least expensive. We also observe that the isolation forest training is constant with respect to the number of clients in the system. 

\section{Conclusion and Future Work}\label{sec:conclusion}

In this paper, we presented \Sys, a novel blockchain-based framework for poison detection in federated learning. Our solution addresses data poisoning attacks from malicious clients by implementing a judge model used to detect anomalies. Model updates that are marked malicious by the judge model are not included in aggregation. We decentralize the role of the trusted central server to clients on the blockchain. The distributed system allows clients to reach consensus on one judge model and verify model updates, without having to rely on a trusted central server. Our implementation of \Sys in the experiments demonstrates its robustness against poisoning attacks and the scalability of producing the judge model.

For future work, \Sys could be expanded to defend against a wider variety of poisoning attacks, making it a more robust framework. The efficiency of reaching consensus on one judge model and the verification process of model updates could be optimized by utilizing quorums of clients instead of requiring all of them to collectively reach consensus. 

\bibliographystyle{IEEEtran.bst}
\bibliography{Citations}   

\end{document}